\documentclass{article}
\usepackage{listings}
\usepackage[hidelinks]{hyperref}
\usepackage{color,rotating,graphicx}
\usepackage[a4paper]{geometry}

\newcount\linecount
\newread\inputlines
\def\wc#1{
\linecount=0
\openin\inputlines=#1
\ifeof\inputlines\else
\catcode`\{=12\catcode`\}=12\readlines\catcode`\{=1\catcode`\}=2\fi
}
\def\readlines{
\read\inputlines to\inputline
\ifeof\inputlines\closein\inputlines
\else\advance\linecount by 1\readlines\fi
}

\def\mytiny#1{\raise.5pt\hbox{\tiny #1}}

\makeatletter 

\def\reducevspace{}
\lst@AddToHook{OnEmptyLine}{\vspace{-.2em}\def\reducevspace{\vspace{-.2em}}}
\lst@AddToHook{EveryPar}{\reducevspace\def\reducevspace{}}

\let\reallst@SkipOrPrintLabel=\lst@SkipOrPrintLabel
\gdef\lst@numberfirstlinetrue{\let\lst@ifnumberfirstline\iftrue}
\def\lst@SkipOrPrintLabel{\ifnum\c@lstnumber=\lst@lastline\lst@numberfirstlinetrue\fi
\reallst@SkipOrPrintLabel}

\lst@AddToHookAtTop{Init}{\def\lst@firstnumber{\lst@firstline}}
\newcount\mylst@lastline
\lst@AddToHook{Init}{\mylst@lastline=\lst@lastline\ifnum\mylst@lastline=9999999\wc{\lstname}\def\lst@lastline{\linecount}\fi}

\DeclareRobustCommand{\mylstname}{%
\begingroup
\def\textendash{-}%
\filename@parse{\lstname}%
\texttt{\filename@base.\filename@ext}%
\endgroup
}

\makeatother

\definecolor{darkgray}{RGB}{102,102,102}
\definecolor{gray}{RGB}{153,153,153}
\definecolor{lightgray}{RGB}{204,204,204}
\definecolor{darkblue}{RGB}{0,0,153}

\def\toolong{$\hspace{-100cm}$}

\begin{document}

\title{A benchmark for C program verification}
\author{%
Marko van Eekelen \and
Daniil Frumin \and
Herman Geuvers \and
L\'eon Gondelman \and
Robbert Krebbers \and
Marc Schoolderman \and
Sjaak Smetsers \and
Freek Verbeek \and
Beno\^it Viguier \and
Freek Wiedijk
}
\date{April 1, 2019}
\maketitle

\begin{abstract}
\noindent
We present twenty-five C programs, as a benchmark for
C program verification using formal methods.
This benchmark can be used for system demonstration,
for comparison of verification effort between systems,
and as a friendly competition.
For this last purpose, we give a scoring formula that allows
a verification system to score up to a hundred points.
\end{abstract}

\section{Introduction}

Here is a challenge for people who are interested in practical C program verification using formal methods.
Show us how to verify the correctness of the twenty-five programs in this paper.
The sources of these programs can be found at:
\begin{center}
\begin{tabular}{l}
\url{https://github.com/cverified/cbench/} \\
\url{https://www.cs.ru.nl/~freek/cbench/}
\end{tabular}
\end{center}
\looseness=-1
If you take this challenge, we will add
your work (or a pointer to it, as you prefer) to the repository.

In our Sovereign project \cite{eek:geu:sme:wie:14}, which has as its focus the safety and correctness of industrial programs, we \emph{are} interested in practical C program verification.
Therefore we would like to have a clearer overview of frameworks for doing this.
At first we planned to only have a few small programs for trying out the different systems (e.g., \cite{app:11}, \cite{cor:cuo:kir:mar:pre:puc:sig:yak:18}, \cite{dah:mos:san:tob:sch:09}, \cite{gre:and:klei:12}, \cite{jac:sma:phi:vog:pen:pie:11}).
However, this then grew into this benchmark.

The original idea of this benchmark was that it should be about five programs.
It will be completely clear what these programs are when one sees the names:
\begin{center}
\begin{tabular}{ll}
\lstinline|fac| & the basis \\
\lstinline|cat| & I/O \\
\lstinline|malloc| & linked datastructures, memory \\
\lstinline|qsort| & arrays, recursion, function pointers \\
\lstinline|sqrt| & floating point
\end{tabular}
\end{center}
The few keywords given here indicate why we chose these five programs.
(Obvious additions to this list are \texttt{fib} and \texttt{gcd}, but we consider these `more of the same'
with respect to \texttt{fac}.)

For each of these programs we now have several variants, ranging from very small with only a few lines of code to significant programs with reasonable behaviors.
We intentionally kept the programs in this benchmark on the
spectrum:
\begin{center}
\begin{picture}(200,35)
\put(0,20){\line(1,0){200}}
\put(0,17){\line(0,1){6}}
\put(200,17){\line(0,1){6}}
\put(0,10){\makebox(0,0){\emph{{\small simplest version}}}}
\put(200,10){\makebox(0,0){\emph{{\small real life implementation}}}}
\end{picture}
\end{center}
\noindent
Altogether this gives twenty-five variations on those five programs.
Three of these are not written by us, but are GNU implementations \cite{sta:02} that are actually used by many people.

To turn this into a benchmark, then, we award four points for a proper verification of any of the programs.
This means that it is possible to score:
$$25 \times \mbox{$4$ points} = \mbox{$100$ points}$$
To get all four points for a program, one should both show that it does not exhibit undefined
behavior and prove its functional correctness.
Points will be subtracted for the following simplifications:
\begin{center}
\begin{tabular}{ll}
no verification of defined behavior & $-1$ \\
no verification of functional correctness & $-1$ \\
verified modified code, still valid C & $-1$ \\
verified only a model of the code & $-2$
\end{tabular}
\end{center}
For example, if one only verifies functional correctness of a model
of the program, one only gets $4 - 1 - 2 = 1$ point ($-1$ for not verifying well-definedness,
and $-2$ for only verifying a model) instead of four.
We include the option to modify the code a bit (with the loss of one point),
because we expect some frameworks to have restrictions on the code they
can verify.

To phrase this more positively, another way to state the way points are awarded is:
\begin{center}
\begin{tabular}{ll}
verification of defined behavior & $+1$ \\
verification of functional correctness & $+1$ \\
verified valid C & $+1$ \\
verified unmodified code & $+1$
\end{tabular}
\end{center}
For the last point it is allowed to add special comments to
the code to enable the verification, like in the ACSL language \cite{bau:cuo:fil:mar:mon:moy:pre:16}.

What it means to correctly verify defined behavior, whether a tool
models C accurately (a subject that can lead to heated debate), and how much modification to the code is allowed
for the program to still be `the same', we intentionally leave open.
This is a friendly challenge, and should not be taken too seriously:
if someone wants to award him- or herself a certain number of points
for a verification effort, they should feel free to do so.
We primarily hope this `benchmark' will be useful for demonstrating and comparing verification frameworks on actual C code.

Exactly what it \emph{means} to `verify' code like this we also intentionally
leave open.
One is free to decide for oneselves what the specification for the programs should be.
In practice, we expect that this means that one has to give:
\begin{itemize}
\item
Restrictions on the C implementation.
\item
Pre-conditions for the code, e.g., restrictions on the input.
\item
Specification of the behavior of the code, e.g., appropriate post-conditions.
\end{itemize}
The C standard \cite{iso:11} has the notion of \emph{implementation-defined
behavior}.
This first item is about making
explicit under what implementations the program should work.

For example, the last programs in our benchmark presuppose IEEE 754-compliant
floating point numbers.
While the object representation of floating point numbers is \emph{not} im\-ple\-men\-ta\-tion-defined (but unspecified),
there is the \lstinline|__STDC_IEC_559__| pre-defined constant, which \emph{is} implementation-defined.
The restriction on the C implementation for these programs then might be that
te implementation has to define that constant appropriately,
in which case we can assume that the object representation for floating point numbers matches the IEEE 754 standard.
Similarly, there might be restrictions on the integer sizes to be `large enough', or struct layouts to be appropriate to the program.

\begin{samepage}
Here then is a list of our twenty-five programs, with a table
of the C features that each of these programs exercises:
\begingroup
\def\u#1{\emph{#1}}
\def\v#1{\begin{sideways}\u{#1}\end{sideways}}
\def\w#1{\hspace{-2pt}\v{#1}\hspace{2pt}}
\def\x{$\times$}
\begin{center}
\medskip
\tabcolsep=2pt
\toolong\begin{tabular}{rclcrcccccccccccccccccccccccccc}
\v{program number} && \v{filename} & \v{type} & \v{lines of code} &$\hspace{9pt}$& \w{function calls} & \w{standard library calls} & \w{other library calls} & \w{recursion} &$\hspace{3pt}$ & \w{preprocessor: includes} & \w{preprocessor: defines} &$\hspace{3pt}$ & \w{input/output} & \w{threads} &$\hspace{3pt}$& \w{exit} & \w{break/continue} & \w{switch} & \w{labels/goto} &$\hspace{3pt}$& \w{arrays/pointers} & \w{function pointers} & \w{structs} & \w{unions} & \w{defined types} & \w{typedefs} &$\hspace{3pt}$& \w{floating point} & \w{IEEE 754 floating point}
\\
\noalign{\vspace{3pt}}
1 && \lstinline|fac1.c| & C & 3 &&&&&&&&&&&&&&&&&&&&&&&&&& \\
2 && \lstinline|fac2.c| & C & 3 &&&&&&&&&&&&&&&&&&&&&&&&&& \\
3 && \lstinline|fac3.c| & F & 9 &&&&&&&&&&&&&&&&&&&&&&&&&& \\
4 && \lstinline|fac4.c| & F & 5 && \x &&& \x &&&&&&&&&&&&&&&&&&&&& \\
5 && \lstinline|fac5.c| & S & 20 &&& \x & \x &&& \x &&& \x &&&&&&&&&&&& \x &&&& \\
6 && \lstinline|fac6.c| & S & 70 && \x & \x && \x && \x & \x && \x &&& \x &&&&&&&&& \x &&&& \\
7 && \lstinline|cat1.c| & C & 2 &&& \x &&&&&&& \x &&&&&&&&&&&&&&&& \\
8 && \lstinline|cat2.c| & S & 13 &&& \x &&&& \x &&& \x &&&&&&&& \x &&&& \x &&&& \\
9 && \lstinline|cat3.c| & S & 76 && \x & \x & \x &&& \x &&& \x &&& \x & \x &&&& \x &&&& \x &&&& \\
10 && \href{https://github.com/coreutils/coreutils/blob/v8.27/src/cat.c}{\lstinline|coreutils/src/cat.c|} & R & ${\sim}800$ && \x & \x & \x &&& \x & \x && \x &&& \x && \x & \x && \x && \x && \x &&&& \\
11 && \lstinline|malloc1.c| & F & 16 &&&&&&&& \x &&&&&&&&&& \x &&&&&&&& \\
12 && \lstinline|malloc2.c| & F & 28 &&&&&&&& \x &&&&&&&&&& \x &&& \x &&&&& \\
13 && \lstinline|malloc3.c| & F & 26 &&&&&&&& \x &&&&&&&&&& \x &&& \x &&&&& \\
14 && \lstinline|malloc4.c| & F & 180 && \x &&&&& \x & \x &&&&&&&&&& \x & \x & \x && \x & \x &&& \\
15 && \href{https://github.com/bminor/glibc/blob/glibc-2.24.90/malloc/malloc.c}{\lstinline|glibc/malloc/malloc.c|} & R & $\hspace{4pt}{\sim}6000$ && \x & \x & \x &&& \x & \x && \x & \x &&&& \x & \x && \x & \x & \x &&&&&& \\
16 && \lstinline|qsort1.c| & F & 23 && \x &&& \x &&&&&&&&&&&&& \x &&&&&&&& \\
17 && \lstinline|qsort2.c| & F & 50 && \x &&& \x &&&&&&&&&&& \x && \x &&&&&&&& \\
18 && \lstinline|qsort3.c| & F & 50 && \x &&& \x &&& \x &&&&&& \x &&&& \x &&&&&&&& \\
19 && \lstinline|qsort4.c| & F & 30 && \x &&& \x && \x & \x &&&&&& \x &&&& \x & \x &&&&&&& \\
20 && \lstinline|qsort5.c| & S & 83 && \x & \x & \x & \x && \x & \x && \x & \x &&& \x &&&& \x &&&& \x &&&& \\
21 && \lstinline|sqrt1.c| & F & 14 &&&&&&&&&&&&&&&&&&&&&&&&& \x & \\
22 && \lstinline|sqrt2.c| & F & 12 &&&&&&&& \x &&&&&&&&&& \x &&&&&&& \x & \x \\
23 && \lstinline|sqrt3.c| & F & 9 &&&&&&&&&&&&&&&&&&&&& \x &&&& \x & \x \\
24 && \lstinline|sqrt4.c| & F & 190 && \x && \x &&&& \x &&&&&& \x & \x &&&&& \x & \x & \x & \x && \x & \x \\
25 && \href{https://github.com/bminor/glibc/blob/glibc-2.24.90/soft-fp/sqrtsf2.c}{\lstinline|glibc/soft-fp/sqrtsf2.c|}$\hspace{8pt}$ & R & ${\sim}500$ && \x && \x &&& \x & \x &&&&&& \x & \x &&&&& \x & \x & \x & \x && \x & \x \\
\end{tabular}\toolong
\medskip
\end{center}
\endgroup
\end{samepage}
\noindent
There are different types of granularity in this list, ranging from only a fragment of a few isolated lines of code
(in which case for example a Hoare triple for that fragment might be proved) to a full multi-file development that spans multiple directories.
The four types are indicated with the following letters:
\begin{center}
\begin{tabular}{ll}
C & code fragment \\
F & one or more functions \\
S & single source file \\
R & real life code, multiple source files
\end{tabular}
\end{center}
The numbers of lines for the GNU programs are only rough indications.
There are \emph{many} files (including many header files) involved in these
programs, and we did not establish exactly which files are needed for
these programs.
(In the list we only list the primary source file, but the verification should
include all other relevant source files.)
Also, the exact line counts will be dependent on the version of the
software.
We do not want to restrict the version that should be used for this
challenge, but while developing this benchmark we looked at version
8.27 of \texttt{coreutils}, and version 2.24.90 of \texttt{glibc}:
\begin{center}
\begin{tabular}{l}
\url{https://github.com/coreutils/coreutils/tree/v8.27} \\
\url{https://github.com/bminor/glibc/tree/glibc-2.24.90}
\end{tabular}
\end{center}
The listings in this paper only show the relevant parts of the programs.
These are the parts that should be verified.
However, in case it is useful to take other parts of these programs into
account, that is of course allowed.
Each of these programs (apart from two of the three GNU ones) is a program that can be compiled on its own,
so each program has a \texttt{main} function in its file too.

\section{Factorial}\label{fac}

\subsection{A simple for loop}
The very first program fragment is a \texttt{for} loop,
multiplying the numbers from 1 to \texttt{n}:
\begin{lstlisting}[linerange={6-8}]
  f = 1;
  for (i = 1; i <= n; i++)
    f = f*i;
\end{lstlisting}
This program has a straight-forward counterpart in most
programming languages, so it should be easy to model it in any
system.
The declarations of \texttt{f}, \texttt{i} and \texttt{n} are not part of this
fragment, but of course should be taken into account when verifying this.

In practice, this program will exhibit integer overflow when \texttt{n} is too large.
With 32-bit integers this happens for $\texttt{n}\ge 13$.
The exact way this is specified in the pre-condition is left open.
For example both $\mbox{\lstinline{n}} < 13$ and $\mbox{\lstinline{n}}! \le \mbox{\lstinline{INT_MAX}}$
might be reasonable choices for this.

\subsection{Turing's program}
The second program fragment is very similar to the previous one, but is a bit more involved.
This is a C rendering of a program by Alan Turing from his 1949 paper \cite{tur:49}
on program verification:
\begin{lstlisting}[linerange={6-8}]
  for (u = r = 1; v = u, r < n; r++)
    for (s = 1; u += v, s++ < r; )
      ;
\end{lstlisting}
This calculates factorial without using multiplication.
The original version of this program was written for the Manchester Mark 1 computer \cite{wie:16}, but as far as we know no C compiler for that computer exists.

\subsection{A function}
We now wrap the code in a function.
This time the verification has to be about pre- and post-conditions
of C functions, not of isolated statements:
\begin{lstlisting}[linerange={1-9},emph={fac,main}]
int
fac(int n)
{
  int f = 1;

  while (n)
    f *= n--;
  return f;
}
\end{lstlisting}
We used more `C-like' idiom in this example.
For example, the update of the counter and the multiplication now
happen together in a single statement, making this example slightly less trivial.

\subsection{Recursion}
We claimed that the for loop has a counterpart in most programming languages,
but this might not hold for functional languages.
Here is the obvious recursive definition of the factorial function in C:
\begin{lstlisting}[linerange={1-5},emph={fac,main}]
int
fac(int n)
{
  return n == 0 ? 1 : n*fac(n - 1);
}
\end{lstlisting}
When called with too large an \texttt{n}
this program can overflow its stack
(for example when compiled with \texttt{gcc} \texttt{-O0} and called with
\texttt{1000000} this happens).
The C standard does not address that issue, but the verification of this
code might take this into account in its pre-condition.

\subsection{Big numbers}
Until now the output of the code is meaningless for most inputs because of overflow.
The program might as well just index in a small array to get the appropriate
output.
Here is a version of the program that uses the GMP library \cite{gra:16} to
compute factorial for larger inputs:
\begin{lstlisting}[emph={main}]
#include <stdio.h>
#include <stdlib.h>
#include "gmp.h"

int
main(int argc, char **argv)
{
  int n, i;
  mpz_t f;

  n = atoi(argv[1]);
  mpz_init(f);
  mpz_set_ui(f, 1);
  for (i = 1; i <= n ; i++)
    mpz_mul_ui(f, f, i);
  mpz_out_str(stdout, 10, f);
  putchar('\n');
  mpz_clear(f);
  return 0;
}
\end{lstlisting}
A specification of this program should also contain appropriate specifications
for the GMP types and functions being used here.
This program computes the factorial of one million (a number of 5,565,709 decimal digits)
in a few minutes.

\subsection{Big numbers without a library}
Here is a simple implementation of factorial with big number arithmetic without
using an external library:
\begin{lstlisting}[emph={calc_fac,print_digits,main}]
#include <stdio.h>
#include <stdlib.h>
#include <stdint.h>

#define M 1000000000
#define D 9

int h = 0;

int
calc_fac(uint32_t *a, int L, uint64_t n)
{
  uint64_t b, c;
  int i, l;

  a[0] = 1;
  for (i = 1; i < L; i++)
    a[i] = 0;
  l = 1;
  while (n) {
    c = 0;
    for (i = 0; i < l || c; i++) {
      b = n*a[i] + c;
      c = b/M;
      a[i] = b - c*M;
    }
    if (i > l)
      l = i;
    n--;
  }
  return l;
}

void
print_digits(int n, int c)
{
  int d;

  if (c <= 0)
    return;
  print_digits(n/10, c - 1);
  d = n%10;
  if (h || d) {
    putchar('0' + d);
    h = 1;
  }
}

int
main(int argc, char **argv)
{
  uint32_t *a;
  int n, m, L, i;

  if (argc != 2) {
    fprintf(stderr, "usage: %s input\n", *argv);
    exit(2);
  }
  n = atoi(argv[1]);
  if (n < 0) {
    fprintf(stderr, "%s: input is negative\n", *argv);
    exit(2);
  }
  L = 0;
  for (m = n; m; m = m/10)
    L++;
  L = L*n/D + 1;
  a = malloc(L*sizeof(*a));
  if (!a) {
    fprintf(stderr, "%s: out of memory\n", *argv);
    exit(1);
  }
  L = calc_fac(a, L, n);
  for (i = L - 1; i >= 0; i--) {
    print_digits(a[i], D);
  }
  putchar('\n');
  return 0;
}
\end{lstlisting}
The output of the digits does not use \texttt{printf}, but prints
the digits individually using \texttt{putchar}.
However, for the error handling \texttt{fprintf} is used.

This program is about ten times slower than the GMP version.

\section{Cat}

\subsection{The simplest input/output program}
Input/output is underrated in C verification.
For example, in the CH$_2$O semantics from Robbert Krebbers' very thorough PhD thesis \cite{kre:15} this
is not properly modelled.
However, we think that input/output (the way that
software interacts with its environment) is a fundamental property of a program
and should be treated as such.

The simplest program with both input and output just copies its input to its output.
In Unix this program is called \texttt{cat}, and the simplest C implementation of this behavior is:
\begin{lstlisting}[linerange={8-9}]
  while ((c = getchar()) != EOF)
    putchar(c);
\end{lstlisting}
How the functional behavior of this program should be {specified} is not addressed in this paper.

\subsection{Buffering}
In practice the previous program is slow, even though \texttt{getchar} internally uses a buffer.
By putting an explicit buffer in the program, it gets as fast as the GNU implementation of \texttt{cat}:
\begin{lstlisting}[emph={main}]
#include <stdio.h>

char buf[128*1024];

int
main()
{
  size_t count;

  while ((count = fread(buf, 1, sizeof(buf), stdin)) > 0)
    fwrite(buf, 1, count, stdout);
  return 0;
}
\end{lstlisting}
This program uses the \lstinline|size_t| type, because that is what the \lstinline|fread| function returns and the \lstinline|fwrite| function expects.
Therefore the verification framework should know about this type, in order to be able to verify this example.

This code is similar to the example program from \cite{gun:myr:kum:nor:17}.

\subsection{A simplified version of the real program}
Here is a simplified version of the GNU implementation of \texttt{cat}.
The main missing functionality is that there are no command line options (`flags').
Also, we simplified the code quite a bit:
\begin{lstlisting}[emph={error,main}]
#include <stdio.h>
#include <string.h>
#include <stdlib.h>
#include <stdint.h>

#include <unistd.h>
#include <fcntl.h>

char *program_name, *infile;
int input_desc, ok = 1;

void
error(int status, int errnum, char *message)
{
  fprintf(stderr, "%s: ", program_name);
  if (errnum)
    perror(message);
  else
    fprintf(stderr, "%s\n", message);
  fflush(stderr);
  if (status)
    exit(status);
  ok = 0;
}

int
main(int argc, char **argv)
{
  size_t insize = 128*1024, page_size = getpagesize(), count, n_rw;
  char *inbuf, *buf, *ptr;
  int argind = 0;

  program_name = argv[argind++];
  infile = "-";
  inbuf = malloc(insize + page_size - 1);
  if (!inbuf)
    error(EXIT_FAILURE, 0, "memory exhausted");
  buf = inbuf + page_size - 1;
  buf = buf - (uintptr_t) buf % page_size;
  do {
    if (argind < argc)
      infile = argv[argind];
    if (*infile == '-' && infile[1])
      error(0, 0, "options not implemented");
    input_desc =
      strcmp(infile, "-") == 0 ? STDIN_FILENO : open(infile, O_RDONLY);
    if (input_desc < 0) {
      error(0, 1, infile);
      continue;
    }
    while (1) {
      count = read(input_desc, buf, insize);
      if (count == (size_t) -1) {
        error(0, 1, infile);
        break;
      }
      if (count == 0)
        break;
      ptr = buf;
      while (count > 0) {
        n_rw = write(STDOUT_FILENO, ptr, count);
        if (n_rw == (size_t) -1)
          break;
        if (n_rw == 0)
          break;
        count -= n_rw;
        ptr += n_rw;
      }
      if (count != 0)
        error(EXIT_FAILURE, 1, "write error");
    }
    if (strcmp(infile, "-") != 0 && close(input_desc) < 0)
      error(0, 1, infile);
  } while (++argind < argc);
  return ok ? EXIT_SUCCESS : EXIT_FAILURE;
}
\end{lstlisting}
This code does not use the standard C library functions
\texttt{fread} and \texttt{fwrite}, but instead
uses the basic Unix system calls \texttt{read} and \texttt{write}.
Therefore, the code includes the \texttt{unistd.h} header file,
and should be verified with a respect to a specification of a Unix
environment.

\subsection{The GNU version}
This is the real program that the previous program was a simplified version of.
When on a Linux system one runs \texttt{cat}, this is the code that is
executed.

The file \texttt{coreutils-8.27/src/cat.c} contains 768 lines of code,
but that code includes eight GNU header files,
links to a function for handling command line options, and uses
the GNU function \texttt{xmalloc}, a wrapper around the C standard library function
\texttt{malloc} which is defined in a different file.
\addtocounter{lstlisting}{1}

\section{Allocate memory}
\subsection{The simplest allocator}
Here is the simplest memory allocator for C possible.
It just cuts off pieces of memory and does not support \texttt{free}:
\begin{lstlisting}[linerange={1-16},emph={malloc}]
#define HEAPSIZE 1000

unsigned char heap[HEAPSIZE];
unsigned char *limit = heap;

unsigned char *
malloc(int n)
{
  unsigned char *ptr;

  if (n <= 0 || n > heap + HEAPSIZE - limit)
    return 0;
  ptr = limit;
  limit += n;
  return ptr;
}
\end{lstlisting}
This code does not satisfy the alignment restrictions on the real
C \texttt{malloc}.
All that is guaranteed is that a block with the appropriate number of bytes is
returned.

This code is very similar to the memory allocator verified in \cite{ben:06,ben:06:1}.

\subsection{A typed allocator}
If one does not need arrays, a memory allocator for C is simple to write.
Here is an allocator that allocates objects of a given type \texttt{T}.
(The definition of this type is intentionally not given in the code
fragment shown, as it should be irrelevant for the verification.)

The code both has an array (to allocate from) and a linked list (for keeping
track of the freed objects), which makes it a nice target for separation logic:
\begin{lstlisting}[linerange={6-33},emph={T_alloc,T_free}]
#define HEAPSIZE 1000

union cell { T allocated; union cell *next_free; } heap[HEAPSIZE];
union cell *limit = heap, *first_free = 0;

T *
T_alloc()
{
  union cell *ptr = first_free;

  if (ptr)
    first_free = first_free->next_free;
  else {
    if (limit >= heap + HEAPSIZE)
      return 0;
    ptr = limit++;
  }
  return &(ptr->allocated);
}

void
T_free(T *T_ptr)
{
  union cell *ptr = (union cell *) T_ptr;

  ptr->next_free = first_free;
  first_free = ptr;
}
\end{lstlisting}

\subsection{A typed allocator that does not need pointer casts}
The previous program needs to get a pointer to a \texttt{union} \texttt{cell}
from a pointer to the allocated \texttt{T} object that is in that cell.
According to the C standard \cite{iso:11} this is allowed, as it says in 6.7.2.1/15:
\begin{quote}
[\,\dots]
A pointer to a union object, suitably converted,
points to each of its members (or if a member is a
bit-field, then to the unit in which it resides), and
vice versa.
[\,\dots]
\end{quote}
However, in some formal versions of the C semantics \cite{kre:16} this is not allowed.
Therefore, here is a variant of this code where the user needs to keep
track of the \texttt{union} \texttt{cell} for him- or herself,
and that problem is not present:
\begin{lstlisting}[linerange={6-31},emph={T_alloc,T_free}]
#define HEAPSIZE 1000

union cell { T allocated; union cell *next_free; } heap[HEAPSIZE];
union cell *limit = heap, *first_free = 0;

union cell *
T_alloc()
{
  union cell *ptr = first_free;

  if (ptr)
    first_free = first_free->next_free;
  else {
    if (limit >= heap + HEAPSIZE)
      return 0;
    ptr = limit++;
  }
  return ptr;
}

void
T_free(union cell *ptr)
{
  ptr->next_free = first_free;
  first_free = ptr;
}
\end{lstlisting}

\subsection{A simplified version of the real program}
Here is a very much simplified version of the GNU implementation of \texttt{malloc} and \texttt{free} in \texttt{glibc}:
\begin{lstlisting}[linerange={7-186},emph={malloc_init_state,free_unsorted_chunk,malloc,free,main}]
#include <stddef.h>

#define HEAPSIZE 1000
union {
  unsigned char bytes[HEAPSIZE];
  max_align_t align;
} heap;

struct malloc_chunk {
  size_t mchunk_prev_size;
  size_t mchunk_size;
  struct malloc_chunk *fd;
  struct malloc_chunk *bk;
  struct malloc_chunk *fd_nextsize;
  struct malloc_chunk *bk_nextsize;
};

typedef struct malloc_chunk *mchunkptr;
typedef struct malloc_chunk *mbinptr;

mchunkptr av_top;
struct malloc_chunk av_bin;

#define chunk2mem(p) ((void*) ((char*) (p) + 2*SIZE_SZ))
#define mem2chunk(mem) ((mchunkptr) ((char*) (mem) - 2*SIZE_SZ))

#define SIZE_SZ (sizeof(size_t))
#define MALLOC_ALIGNMENT (2*SIZE_SZ < _Alignof(max_align_t) \
  ? _Alignof(long double) : 2*SIZE_SZ)
#define MALLOC_ALIGN_MASK (MALLOC_ALIGNMENT - 1)

#define MIN_CHUNK_SIZE (sizeof(struct malloc_chunk))
#define MINSIZE (unsigned long) \
  (((MIN_CHUNK_SIZE + MALLOC_ALIGN_MASK) & ~MALLOC_ALIGN_MASK))

#define PREV_INUSE 0x1
#define SIZE_BITS (PREV_INUSE)

#define chunksize(p) (chunksize_nomask(p) & ~(SIZE_BITS))
#define prev_inuse(p) ((p)->mchunk_size & PREV_INUSE)
#define chunksize_nomask(p) ((p)->mchunk_size)
#define prev_size(p) ((p)->mchunk_prev_size)
#define chunk_at_offset(p, s) ((mchunkptr) (((char *) (p)) + (s)))

#define set_head(p, s) ((p)->mchunk_size = (s))
#define set_foot(p, s) \
  (((mchunkptr) ((char *) (p) + (s)))->mchunk_prev_size = (s))

#define inuse_bit_at_offset(p, s) \
  (((mchunkptr) (((char *) (p)) + (s)))->mchunk_size & PREV_INUSE)
#define set_inuse_bit_at_offset(p, s) \
  (((mchunkptr) (((char *) (p)) + (s)))->mchunk_size |= PREV_INUSE)
#define clear_inuse_bit_at_offset(p, s) \
  (((mchunkptr) (((char *) (p)) + (s)))->mchunk_size &= ~PREV_INUSE)

#define unlink(P, BK, FD) \
  { FD = P->fd; BK = P->bk; FD->bk = BK; BK->fd = FD; }

static void
free_unsorted_chunk(mchunkptr p, size_t size)
{
  mchunkptr fwd, bck;

  set_head(p, size | PREV_INUSE);
  set_foot(p, size);
  bck = &av_bin;
  fwd = bck->fd;
  if (fwd != bck) {
    size |= PREV_INUSE;
    while ((unsigned long) size < chunksize_nomask(fwd))
      fwd = fwd->fd_nextsize;
    if ((unsigned long) size ==
        (unsigned long) chunksize_nomask(fwd))
    {
      p->fd_nextsize = p->bk_nextsize = NULL;
      fwd = fwd->fd;
    } else {
      p->fd_nextsize = fwd;
      p->bk_nextsize = fwd->bk_nextsize;
      fwd->bk_nextsize = p;
      p->bk_nextsize->fd_nextsize = p;
    }
    bck = fwd->bk;
  } else
    p->fd_nextsize = p->bk_nextsize = p;
  p->bk = bck;
  p->fd = fwd;
  fwd->bk = p;
  bck->fd = p;
}

static void
malloc_init_state()
{
  size_t size;

  av_bin.fd = av_bin.bk = &av_bin;
  size = (HEAPSIZE - MIN_CHUNK_SIZE + 2*SIZE_SZ) & ~MALLOC_ALIGN_MASK;
  size -= MALLOC_ALIGNMENT;
  av_top = (mchunkptr) (heap.bytes + MALLOC_ALIGNMENT - 2*SIZE_SZ);
  set_inuse_bit_at_offset(av_top, size);
  free_unsorted_chunk(av_top, size);
}

void *
malloc(size_t bytes)
{
  size_t nb, size;
  mbinptr bin;
  mchunkptr victim, remainder, fwd, bck;
  unsigned long remainder_size;

  nb = bytes + SIZE_SZ + MALLOC_ALIGN_MASK;
  nb = nb < MINSIZE ? MINSIZE : nb & ~MALLOC_ALIGN_MASK;
  if (!(bin = &av_bin))
    malloc_init_state();
  if ((victim = bin->fd) != bin
      && (unsigned long) chunksize_nomask(victim)
         >= (unsigned long) (nb)) {
    victim = victim->bk_nextsize;
    while (((unsigned long) (chunksize(victim))
           < (unsigned long) (nb)))
      victim = victim->bk_nextsize;
    if (victim != bin->bk
        && chunksize_nomask(victim) == chunksize_nomask(victim->fd))
      victim = victim->fd;
    size = chunksize(victim);
    remainder_size = size - nb;
    unlink(victim, bck, fwd);
    if (remainder_size < MINSIZE)
      set_inuse_bit_at_offset(victim, size);
    else {
      remainder = chunk_at_offset(victim, nb);
      set_head(victim, nb | PREV_INUSE);
      free_unsorted_chunk(remainder, remainder_size);
    }
    return chunk2mem(victim);
  }
  victim = av_top;
  size = chunksize(victim);
  if ((unsigned long) (size) >= (unsigned long) (nb + MINSIZE)) {
    remainder_size = size - nb;
    remainder = chunk_at_offset(victim, nb);
    av_top = remainder;
    set_head(victim, nb | PREV_INUSE);
    set_head(remainder, remainder_size | PREV_INUSE);
    return chunk2mem(victim);
  }
  return 0;
}

void
free(void *mem)
{
  mchunkptr p, nextchunk, bck, fwd;
  size_t size, nextsize, prevsize;

  p = mem2chunk(mem);
  size = chunksize(p);
  nextchunk = chunk_at_offset(p, size);
  nextsize = chunksize(nextchunk);
  if (!prev_inuse(p)) {
    prevsize = prev_size(p);
    size += prevsize;
    p = chunk_at_offset(p, -((long) prevsize));
    unlink(p, bck, fwd);
  }
  if (nextchunk != av_top) {
    if (!inuse_bit_at_offset(nextchunk, nextsize)) {
      unlink(nextchunk, bck, fwd);
      size += nextsize;
    } else
      clear_inuse_bit_at_offset(nextchunk, 0);
    free_unsorted_chunk(p, size);
  } else {
    size += nextsize;
    set_head(p, size | PREV_INUSE);
    av_top = p;
  }
}
\end{lstlisting}
Some features of the real GNU malloc that are missing from this version:
\begin{itemize}
\item no fastbins
\item just a single regular bin
\item no bitmap to quickly find non-empty bins (not needed, as there is just one bin)
\item no \texttt{sbrk} to get more memory (just one static array)
\item memory never returned to the system (the \lstinline|av_top| pointer never decreases)
\item no memory mapped blocks
\item no thread-awareness
\end{itemize}
However, the free list in the one bin is still kept in sorted order, and still uses
the \lstinline|fd_nextsize| and \lstinline|bk_nextsize| pointers to search
for an appropriate block.

\subsection{The GNU version}
And again, the last program of this section is the full GNU implementation of \texttt{malloc} and \texttt{free} in \texttt{glibc}.
The main file \lstinline|glibc/malloc/malloc.c| in version 2.24.90 of \texttt{glibc} has 5,289 lines of code, but this file depends on many other files.

The file contains the sentence:
\begin{quote}
The implementation is in straight, hand-tuned ANSI C.
\end{quote}
Exactly what it means to say that this code is `ANSI C', and whether this is actually true, is an interesting question.
\addtocounter{lstlisting}{1}

\section{Quicksort}
\subsection{Straight-forward quicksort}\label{qsort1}
Here is a very straight-forward version of a quicksort function for
an array of doubles:
\begin{lstlisting}[linerange={4-26},emph={swap,quicksort,main}]
double a[N];

void
quicksort(int m, int n)
{
  int i, j;
  double pivot, tmp;

  if (m < n) {
    pivot = a[n];
    i = m; j = n;
    while (i <= j) {
      while (a[i] < pivot) i++;
      while (a[j] > pivot) j--;
      if (i <= j) {
        tmp = a[i]; a[i] = a[j]; a[j] = tmp;
        i++; j--;
      }
    }
    quicksort(m, j);
    quicksort(i, n);
  }
}
\end{lstlisting}
This function of course has to be called with \lstinline|quicksort(0, N - 1)|.

We tried to make this code as simple as possible.
For example it contains no \lstinline|return| or \lstinline|break| and only has a single swap.

There is a variant of this in which there only is a single comparison between
\lstinline|i| and \lstinline|j| (with the bounds in the recursive
call different), but that really makes use of the fact
that we used \lstinline|a[n]| for the pivot.
For that reason we decided not to take that variant here, even though
it is still shorter.
The version here also works with different choices for the pivot,
like taking the median or mean of some elements in the array.

This program has the property that if the array is already sorted, then it needs stack space for the recursion proportional to the length of the array.
On a Linux system with a (default) \texttt{stacksize} of 8MB, the full program, which
has \texttt{N} equal to 666,666, fails after about three minutes with the error:
\begin{quote}
\lstinline|segmentation fault|
\end{quote}
It would be nice if the verification of this program would be compatible with this behavior, i.e., that it would not be possible to prove this program to be correct for too large \texttt{N}.

\subsection{Hoare's version of quicksort}
Here is a variation on the previous program.
This is a C rendering of the programs in the original publication on
quicksort by Tony Hoare
\cite{hoa:61}:
\begin{lstlisting}[linerange={17-66},emph={exchange,random,partition,quicksort,main}]
void
partition(double *a, int m, int n, int *_i, int *_j)
{
  double x;
  int f, i, j;

  f = random(m, n);
  x = a[f];
  i = m;
  j = n;
up:
  for (i = i; i <= n; i++)
    if (x < a[i])
      goto down;
  i = n;
down:
  for (j = j; j >= m; j--)
    if (a[j] < x)
      goto change;
  j = m;
change:
  if (i < j) {
    exchange(&a[i], &a[j]);
    i = i + 1;
    j = j - 1;
    goto up;
  } else
  if (i < f) {
    exchange(&a[i], &a[f]);
    i = i + 1;
  } else
  if (f < j) {
    exchange(&a[f], &a[j]);
    j = j - 1;
  }
  *_i = i;
  *_j = j;
}

void
quicksort(double *a, int m, int n)
{
  int i, j;

  if (m < n) {
    partition(a, m, n, &i, &j);
    quicksort(a, m, j);
    quicksort(a, i, n);
  }
}
\end{lstlisting}
This program uses \lstinline{goto} instead of \lstinline{break}.
The original Algol 60 version of this code uses the call-by-name feature of Algol 60 (which is
more like macro expansion),
which is modeled here by giving the \lstinline{partition} function pointers
to the variables to be modified.

\subsection{The algorithm in the real program}
If we take the GNU implementation of quicksort, and simplify it and also make it
only apply to arrays of doubles, this is what one gets:
\begin{lstlisting}[linerange={1-50},emph={SWAP,quicksort,main}]
#define SWAP(a, b) do { tmp = *a; *a = *b; *b = tmp; } while(0)

void
quicksort(double *base_ptr, int total_elems)
{
  double tmp, *lo, *hi, *left_ptr, *right_ptr, *mid;

  if (total_elems == 0)
    return;
  lo = base_ptr;
  hi = lo + total_elems - 1;
  while (lo < hi) {
    mid = lo + ((hi - lo) >> 1);
    if (*mid < *lo)
      SWAP(mid, lo);
    if (*hi < *mid) {
      SWAP(mid, hi);
      if (*mid < *lo)
        SWAP(mid, lo);
    }
    left_ptr = lo + 1;
    right_ptr = hi - 1;
    do {
      while (*left_ptr < *mid)
        left_ptr++;
      while (*mid < *right_ptr)
        right_ptr--;
      if (left_ptr < right_ptr) {
        SWAP(left_ptr, right_ptr);
        if (mid == left_ptr)
          mid = right_ptr;
        else if (mid == right_ptr)
          mid = left_ptr;
        left_ptr++;
        right_ptr--;
      } else if (left_ptr == right_ptr) {
        left_ptr++;
        right_ptr--;
        break;
      }
    } while (left_ptr <= right_ptr);
    if (right_ptr - lo > hi - left_ptr) {
      quicksort(left_ptr, hi - left_ptr + 1);
      hi = right_ptr;
    } else {
      quicksort(lo, right_ptr - lo + 1);
      lo = left_ptr;
    }
  }
}
\end{lstlisting}
This is just a variation on the previous programs, but it has
a few nice features, like using a macro for swapping elements in the array.
One probably wants to prove this \emph{macro} correct,
which means that just verifying the macro-expanded version of the
program will not be too attractive in this case.

Apart from the fact that this function only sorts doubles instead
of sorting arrays with elements of an arbitrary type,
it omits two optimizations from the GNU version of this code.
This uses insertion sort when the size of the subarray drops
below a certain threshold,
and it manages a (short) \emph{explicit} stack for the recursive calls,
instead of using the recursion of C.

\subsection{Quicksort on arbitrary types}
Here is a version of the simple version of quicksort that
we gave in \ref{qsort1}, but this time for arbitrary types, so this is an
implementation of the interface of the \lstinline{qsort} function in the C standard library:
\begin{lstlisting}[linerange={1-30},emph={swap,qsort,main}]
#include <string.h>

#define a(i) (((unsigned char *) base) + i*size)

void
qsort(void *base, size_t nmemb, size_t size,
  int (*compar)(const void *, const void *))
{
  int m, n, i, j;
  unsigned char pivot[size], tmp[size];

  m = 0;
  n = nmemb - 1;
  if (m < n) {
    memcpy(pivot, a(n), size);
    i = m; j = n;
    while (i <= j) {
      while ((*compar)(a(i), pivot) < 0) i++;
      while ((*compar)(a(j), pivot) > 0) j--;
      if (i <= j) {
        memcpy(tmp, a(i), size);
        memcpy(a(i), a(j), size);
        memcpy(a(j), tmp, size);
        i++; j--;
      }
    }
    qsort(a(m), j - m + 1, size, compar);
    qsort(a(i), n - i + 1, size, compar);
  }
}
\end{lstlisting}
Notable features of this program are function pointers,
\lstinline|(void *)| pointers, and the use of \lstinline|memcpy| to
copy C objects.

\subsection{Parallel quicksort}
Finally, here is a version of quicksort that uses threads.
The basic idea of this program is that at most \lstinline|THREADS| threads will be
running.
Once a recursive call gets a piece of the array of which the size is
below \lstinline|THRESHOLD| (which
means the data will fit in the cache of one of the cores), a separate
thread will be used for this if one is available.
The count of free threads is kept in the variable \lstinline|available|
for which a mutex is used when updating it.

We did not use C11 threads for this, because we did not have a good implementation of it available.
Instead this uses the pthreads library.
The means the program might need to be compiled
with \lstinline{-lpthread}.

Again, algorithmically this is just the simple algorithm from \ref{qsort1}:
\begin{lstlisting}[emph={run_quicksort,quicksort,main}]
#include <stdio.h>
#include <stdlib.h>
#include <pthread.h>

#define N 666666
double a[N];

#define THRESHOLD 0x2000
#define THREADS 16

pthread_mutex_t mutex;
int active = 1;

void quicksort(int, int, int);
struct quicksort_args {int m; int n;};

void *
run_quicksort(void *void_args)
{
  struct quicksort_args *args = void_args;

  quicksort(args->m, args->n, 1);
  return 0;
}

void
quicksort(int m, int n, int threaded)
{
  int i, j;
  double pivot, tmp;
  pthread_t thread;
  struct quicksort_args args;

  if (m < n) {
    pivot = a[n];
    i = m; j = n;
    while (i <= j) {
      while (a[i] < pivot) i++;
      while (a[j] > pivot) j--;
      if (i <= j) {
        tmp = a[i]; a[i] = a[j]; a[j] = tmp;
        i++; j--;
      }
    }
    if (threaded) {
      if (n - m < THRESHOLD)
        threaded = 0;
      else {
        pthread_mutex_lock(&mutex);
        if (active < THREADS) {
          active++;
          pthread_mutex_unlock(&mutex);
          args.m = m;
          args.n = j;
          pthread_create(&thread, 0, &run_quicksort, &args);
          quicksort(i, n, 1);
          pthread_mutex_lock(&mutex);
          active--;
          pthread_mutex_unlock(&mutex);
          pthread_join(thread, 0);
          return;
        }
        pthread_mutex_unlock(&mutex);
      }
    }
    quicksort(m, j, threaded);
    quicksort(i, n, threaded);
  }
}

int
main()
{
  int i;

  pthread_mutex_init(&mutex, 0);
  for (i = 0; i < N; i++)
    a[i] = random();
  quicksort(0, N - 1, 1);
  for (i = 0; i < N; i++)
    printf("%f\n", a[i]);
  return 0;
}
\end{lstlisting}

\section{Square root}
\subsection{Newton's approximation}
The last class of examples in this benchmark are implementations
of the square root function on single precision floating point numbers.
Here is a simple version of an implementation that uses the Newton method:
\begin{lstlisting}[linerange={1-14},emph={sqrt_newton,main}]
float
sqrt_newton(float x)
{
  float y, z;

  if (x <= 0)
    return 0;
  y = x >= 1 ? x : 1;
  do {
    z = y;
    y = (z + x/z)/2;
  } while (y < z);
  return y;
}
\end{lstlisting}
This will not always give exactly the right answer according to the IEEE 754
standard, but it will be close.
If a \lstinline|float| has finitely many different representations
(which one might expect the C standard would enforce, because there are only
finitely many different object representations), then this program will
always terminate.

It is interesting exactly what is provable about this program
in a C implementation where one does not know how the floating
point implementation works in detail.
Maybe if there are some guarantees about relative error bounds
of the arithmetic operations, an error bound on the result of this
function can be proved.

\subsection{An approximation of square root by bit shifting}\label{bitshift}
The IEEE 754 representation of floating point numbers allows
for an approximation to the square root function
that can be calculated extremely fast.
The counterpart for the inverse square root function (both square
root and inverse square root
use exactly the same trick) is known
as `the \lstinline|0x5f3759df| method',
after the magic constant that appears in it.
(Comments in the first program that had this constant
in it were \emph{evil floating point bit level hacking} and \emph{what the fuck?})

\begin{figure}
\begin{center}
\includegraphics[width=.9\textwidth]{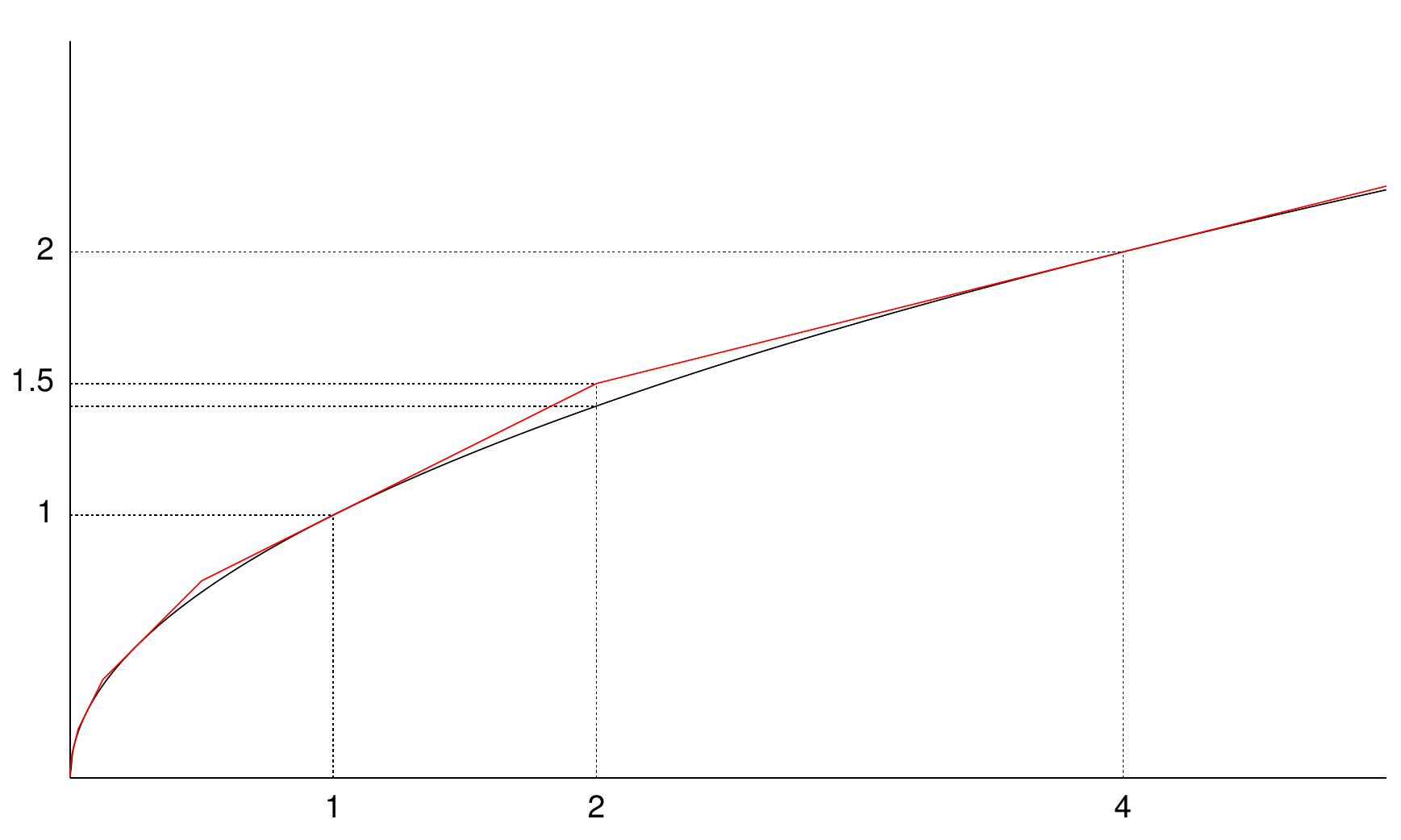}
\end{center}
\caption{The graph of \lstinline|sqrt_approx|.}\label{sqrt_approx}
\end{figure}

The implementation just consists of considering the bits of the number by casting it to an integer,
shifting those bits one place to the right, and then adding a magic constant to that integer representation:
\begin{lstlisting}[linerange={1-12},emph={sqrt_approx,main}]
#define BITS(x) (*((unsigned *) &x))
#define FLOAT(x) (*((float *) &x))

float
sqrt_approx(float x)
{
  float one = 1;
  unsigned i;

  i = (BITS(x) >> 1) + BITS(one) - (BITS(one) >> 1);
  return FLOAT(i);
}
\end{lstlisting}
It is surprising that this gives a sensible result.
One is shifting one of the bits of the exponent into a bit of the
mantissa, how is that going to make sense?
However, when analyzing it more closely \cite{bli:97}, it turns
out that this \emph{does} make sense, and that for sufficiently large inputs, it gives a continuous piecewise linear function that 
approximates the square root function quite well (it is exact
on powers of four, see the graph in Fig.~\ref{sqrt_approx}.)
The maximum error occurs when the argument is twice a power of four, and the
relative error on those values is $(1\frac{1}{2} - \sqrt{2})/\sqrt{2} = \frac{3}{4}\sqrt{2} - 1 \approx 6.0660\% \approx 2^{-4.0431}$,
which means that generally the first three bits of the fractional part will be correct.

Unlike the code in the next section, this function also makes sense if \lstinline|float|s
do not have 32 bits single precision.
All that is needed is that the bits are laid out in the style of a
IEEE 754 representation.
As was noted before, this can be established by the implementation-defined
property of \lstinline|__STDC_IEC_559__| having an appropriate definition.

We had some discussion whether to include this program in our benchmark.
We wanted our examples to be `realistic', and this seems more of
a cute hack than something resembling realistic code.
Also, nowadays square root is supported by hardware, and getting a good approximation
as a start for Newton's method is not really important anymore.
However, this program is about the bit representation of floating point
numbers, and about using casts to reinterpret one type as another (which is
a common thing in C).
For this reason we decided to still include this
as one of our examples.

\subsection{An approximation of square root by bit shifting that does not violate effective types}\label{sqrt3}
The previous program uses type casts to get at the bits of a floating point number,
in order to be able to shift them.
However, that violates the \emph{effective type} rules, as described
in 6.5/7 of the C11 standard.
In fact, when compiled using GCC with the \lstinline|-Wall| and \lstinline|-O3| flags,
the compiler emits the warning:
\begin{quote}
\lstinline[language=]|warning: dereferencing type-punned pointer will break strict-|\\
\lstinline[language=]|aliasing rules [-Wstrict-aliasing]|
\end{quote}
We now give a version of the same program that uses a \lstinline|union| to
get at the bits of the \lstinline|float|, which means it
does not suffer from this problem:
\begin{lstlisting}[linerange={1-9},emph={sqrt_approx,main}]
float
sqrt_approx(float x)
{
  union { float x; unsigned i; } u;

  u.x = x;
  u.i = (u.i >> 1) + 0x1fc00000;
  return u.x;
}
\end{lstlisting}
Using a \lstinline|union| to get at the bits of an object is allowed by the C11 standard, as in
footnote 95 of 6.5.2.3/3 it reads:
\begin{quote}
If the member used to read the contents of a union object is not the same as the member last used to
store a value in the object, the appropriate part of the object representation of the value is reinterpreted
as an object representation in the new type as described in 6.2.6 (a process sometimes called ``type
punning''). This might be a trap representation.
\end{quote}
When using \lstinline|union|s, it is harder to write the constant to be
added to the expression in a way that it will work whatever
the number of bits in the floating point representation, and
in such a way that the compiler will insert it as a constant in the code
at compile time.
To keep the program small, we therefore expanded this constant to its value for 32-bit
floating point numbers: \lstinline|0x1fc00000|. 
This means that to work correctly, this program needs an implementation with 32-bit IEEE 754 \lstinline|float|s.

\subsection{A simplified version of a serious implementation}
The GNU library \texttt{glibc} contains software implementations of
the IEEE 754 operations in the subdirectory \texttt{soft-fp}.
In the case of square root, the algorithm used is the square root
analog of long division.
The GNU code for this consists of a whole stack of macro definitions.
When expanding all these definitions, and simplifying everything quite a bit, one gets:
\begin{lstlisting}[linerange={1-190},emph={__sqrtsf2,main}]
typedef float SFtype;

#define FRACBITS                24
#define EXPBITS                 8
#define EXPBIAS                 127
#define EXPMAX                  255

#define W_TYPE                  unsigned long
#define I_TYPE                  long
#define W_TYPE_SIZE             32

#define CLS_NORMAL              0
#define CLS_ZERO                1
#define CLS_INF                 2
#define CLS_NAN                 3

union UNION
{
  SFtype flt;
  struct STRUCT_LAYOUT
  {
    unsigned frac:FRACBITS - 1;
    unsigned exp:EXPBITS;
    unsigned sign:1;
  } bits;
};

SFtype
__sqrtsf2(SFtype a)
{
  SFtype r;
  union UNION RAW_flo;
  I_TYPE shift;
  I_TYPE A_c;
  I_TYPE A_s;
  I_TYPE A_e;
  W_TYPE A_f;
  I_TYPE R_c;
  I_TYPE R_s;
  I_TYPE R_e;
  W_TYPE R_f;
  W_TYPE S_f;
  W_TYPE T_f;
  W_TYPE q;

  RAW_flo.flt = a;
  A_s = RAW_flo.bits.sign;
  A_e = RAW_flo.bits.exp;
  A_f = RAW_flo.bits.frac;
  switch (A_e) {
  default:
    A_c = CLS_NORMAL;
    A_f |= ((W_TYPE) 1 << (FRACBITS - 1));
    A_f <<= 3;
    A_e -= EXPBIAS;
    break;
  case 0:
    if (A_f == 0)
      A_c = CLS_ZERO;
    else {
      shift = __builtin_clzl(A_f);
      shift -= (W_TYPE_SIZE - FRACBITS);
      A_c = CLS_NORMAL;
      A_f <<= (shift + 3);
      A_e -= (EXPBIAS - 1 + shift);
    }
    break;
  case EXPMAX:
    if (A_f == 0)
      A_c = CLS_INF;
    else
      A_c = CLS_NAN;
    break;
  }
  switch (A_c) {
  case CLS_NAN:
    R_c = CLS_NAN;
    R_s = A_s;
    R_f = A_f;
    break;
  case CLS_INF:
    if (A_s) {
      R_c = CLS_NAN;
      R_s = 1;
      R_f = ((W_TYPE) 1 << (FRACBITS - 2));
    } else {
      R_c = CLS_INF;
      R_s = 0;
    }
    break;
  case CLS_ZERO:
    R_c = CLS_ZERO;
    R_s = A_s;
    break;
  case CLS_NORMAL:
    if (A_s) {
      R_c = CLS_NAN;
      R_s = 1;
      R_f = ((W_TYPE) 1 << (FRACBITS - 2));
      break;
    }

    R_c = CLS_NORMAL;
    R_s = 0;
    R_e = A_e >> 1;
    if (A_e & 1)
      A_f += A_f;
    S_f = 0;
    R_f = 0;
    q = ((W_TYPE) 1 << (3 + FRACBITS)) >> 1;
    while (q != ((W_TYPE) 1 << 2)) {
      T_f = S_f + q;
      if (T_f <= A_f) {
        S_f = T_f + q;
        A_f -= T_f;
        R_f += q;
      }
      A_f += A_f;
      q >>= 1;
    }
    if (A_f) {
      if (S_f < A_f)
        R_f |= ((W_TYPE) 1 << 2);
      R_f |= ((W_TYPE) 1 << 0);
    }

  }
  switch (R_c) {
  case CLS_NORMAL:
    R_e += EXPBIAS;
    if (R_e > 0) {
      if (R_f & 7) {
        if ((R_f & 15) != ((W_TYPE) 1 << 2))
          R_f += ((W_TYPE) 1 << 2);
      }
      if (R_f & ((W_TYPE) 1 << (3 + FRACBITS))) {
        R_f &= ~((W_TYPE) 1 << (3 + FRACBITS));
        R_e++;
      }
      R_f >>= 3;
      if (R_e >= EXPMAX) {
        R_c = CLS_INF;
        R_e = EXPMAX;
        R_f = 0;
      }
    } else {
      R_e = -R_e + 1;
      if (R_e <= (3 + FRACBITS)) {
        R_f = (R_f >> R_e);
        if (R_f & 7) {
          if ((R_f & 15) != ((W_TYPE) 1 << 2))
            R_f += ((W_TYPE) 1 << 2);
        }
        if (R_f & (((W_TYPE) 1 << (3 + FRACBITS)) >> 1)) {
          R_e = 1;
          R_f = 0;
        } else {
          R_e = 0;
          R_f >>= 3;
        }
      } else {
        R_e = 0;
        if (R_f != 0) {
          R_f = 1;
          if ((R_f & 7) && ((R_f & 15) != ((W_TYPE) 1 << 2)))
            R_f += ((W_TYPE) 1 << 2);
          R_f >>= 3;
        }
      }
    }
    break;
  case CLS_ZERO:
    R_e = 0;
    R_f = 0;
    break;
  case CLS_INF:
    R_e = EXPMAX;
    R_f = 0;
    break;
  case CLS_NAN:
    R_e = EXPMAX;
    R_f |= ((W_TYPE) 1 << (FRACBITS - 2));
    break;
  }
  RAW_flo.bits.frac = R_f;
  RAW_flo.bits.exp = R_e;
  RAW_flo.bits.sign = R_s;
  r = RAW_flo.flt;
  return r;
}
\end{lstlisting}
This code implements IEEE 754 behavior, but (as the primary simplification)
only implements the default rounding mode.
The full GNU version supports the other rounding modes as well.
Also, the full version implements more possible variant behaviors on not-a-numbers.

Most of this code is about getting at the bits at the start and reassembling the
floating point number at the end, and about
handling various special cases like infinities, not-a-numbers, and
denormalized numbers.
The actual algorithm for the `long division' analog
is in lines 103--125.

\subsection{The GNU version}
And the last program in our verification benchmark is the actual implementation
of square root in \texttt{libc/soft-fp}.
That code is a bit obfuscated, because many of the macros implement the operations
for different floating point sizes at the same time.
The function that should be verified is \lstinline|__sqrtsf2|,
implemented in the file \lstinline|sqrtsf2.c|:
\begin{lstlisting}[linerange={31-49},emph={__sqrtsf2}]
#include "soft-fp.h"
#include "single.h"

SFtype
__sqrtsf2 (SFtype a)
{
  FP_DECL_EX;
  FP_DECL_S (A);
  FP_DECL_S (R);
  SFtype r;

  FP_INIT_ROUNDMODE;
  FP_UNPACK_S (A, a);
  FP_SQRT_S (R, A);
  FP_PACK_S (r, R);
  FP_HANDLE_EXCEPTIONS;

  return r;
}
\end{lstlisting}
Of course the macros defined through the \lstinline|#include| lines pull in lots of other code from other files, and expand this program significantly.

In the latest versions of \texttt{glibc} the source code of square root in \texttt{libc/soft-sp} has been moved to a larger file that contains many arithmetical operations together.
However, we prefer to have the 25th program in our benchmark be about a single small source file that we essentially can include in this paper.
For this reason, this benchmark stays linked to version 2.24.90 of \texttt{glibc}.

\section{Versions}

The benchmark presented in this paper will go through various versions.
We will use dates for versioning.
The initial version is dated April 1, 2019, and might be
referred to as the \emph{April fool's} version.

Every time a problem with one of the programs in this
benchmark is discovered, we will modify the program to remove the problem, and give
the benchmark a new version date.
In that case we will add a short description to this section of the problem, how it was discovered and how the program was changed to remove the problem.
This means that this section will be some kind of high level changelog.

Note that bad behavior outside the
intended domain of behavior of a program does not count as a problem.
For example, integer overflow in the earlier factorial programs in Section~\ref{fac}
is not a problem.
This should be handled by a precondition of the correctness statement of the code, and
does not merit modification of the code in the benchmark.

\section{Discussion}
We can imagine different reactions to the collection of programs that we present here:
\begin{itemize}
\item
These programs are too easy.

\item
These programs are too hard.

\item
These programs are not too easy or too hard, but not interesting.

\end{itemize}
\noindent
In the first case we would like to see how to verify them.
If they are easy, showing how to do that should not be much work, right?
In the second case we would like to know how far the technology
currently reaches.
We intentionally provided a spectrum from very small fragments
to larger pieces of code, so if you have the second reaction then just show us how to do the easier ones.

However, in the third case we disagree.
This is based on actual programs from the GNU project,
and as such are representative of real C programming.
Why is it not interesting to be able to verify what
to us are characteristic C programs?

It was tempting to also include programs in our benchmark that are cute and/or dazzling.
(The square root approximation by bit shifting in Section~\ref{bitshift} was borderline in this respect.)
For instance, consider two possible additions to the factorial
section,
the first being a surprising application of the preprocessor:
\begin{lstlisting}
#include <stdio.h>

/* after all the macro violence, PRIM(fac) will be left in the output;
   one solution is to define these as ... existing nowhere */

extern int fac();
extern int (*PRIM())();

#define NOTHING
#define PRIM(f) f##_prim

/* NOTHING between PRIM and (f) prevens expansion of PRIM */
#define REC PRIM NOTHING

/* this simply forces the re-scanning of x (for macros) many times */
#define EVAL1(x) x
#define EVAL2(x) EVAL1(EVAL1(EVAL1(EVAL1(x))))
#define EVAL3(x) EVAL2(EVAL2(EVAL2(EVAL2(x))))

#define fac_prim(n) (n)<=0? 1 : (n)*(REC(fac)(n-1))
#define fac(n) EVAL3(PRIM(fac)(n))

int main()
{
    printf("%ld\n", fac(20L));
}
\end{lstlisting}
\begin{lstlisting}[language=]
   #include<stdarg.h>/*Y=\f*/
  #include<stdio.h>/*.(\x.f(*/
  #define z return/*xx))(\x.*/
  #include<stdlib.h>/*f(xx))*/
   typedef union w w;union w{
   #define L(f,y,x,t) w _ ##\
   f(w x,...){va_list argh;w\
   y; va_start(argh,x);y= *\
   (va_arg(argh,w*));va_end\
   (argh);z(t);} w f (w fw)\
   {r((w){.c= &_## f}, fw);}
   int _; void*p;w(*c)(w,...
   #define r(x,y)w*k=malloc\
    (sizeof(w)<<1); k[0]= x\
    ; k[1]=y; z((w) {.p= k})
    #define l(f,x,t)w _##f(\
    w x,...){z(t);}w f={.p\
    =(w[1]){{.c= &_##f}}};
    );};w a(w f,w x){w*d=f
     .p;z((*d->c)(x,d+1)


             ) ; }
      w a_(w f,w fw){r(f,
   fw);}w _(w f){w*k=f.p,r=
  a(*k,k[1]); free(k) ;z(r);}
 L(F,f,x,(w){._=x._?x._*a(_(f),
(w){._=x._-1})._:1})l(F_,f,F(f)
)L(  W,(f),/*=\*/x,a(f,a_(x,x))
)l(      Y,/*=\*/f,a(W(f),W(f))
 )int main(){printf("%.""f\n",
    a(a(Y,F_),(w){._=10})._
        /60./60/24/*d */
             ) ; }
\end{lstlisting}
(it turns out that the factorial is calculated here at compile time),
and the second being a price winning submission to the international
obfuscated C code contest
(it implements a functional language and then does the recursion by
evaluating Church's $Y$-combinator.)
However, we do not consider these two programs representative of
real life C programming, so we resisted the temptation to include them in our benchmark.

There are twenty-five programs in this paper, and there are multiple
frameworks for formal C verification out there, like the ones described
in \cite{app:11}, \cite{cor:cuo:kir:mar:pre:puc:sig:yak:18}, \cite{dah:mos:san:tob:sch:09}, \cite{gre:and:klei:12} and \cite{jac:sma:phi:vog:pen:pie:11}.
We think this could be a great source for student projects.
Just say, for example: `prove the correctness of program number three in framework
number one' (actually something close to that one already has been done, see the slides of \cite{wie:16:1}).
That way we will get a good comparison of what are the strengths and
weaknesses of the different current approaches for formal C verification.

\section{Related work}

Benchmarks/competitions very similar to the one presented in this paper
already exist:
\begin{itemize}
\item
VACID-0 \cite{lei:mos:10}: \\
\url{https://archive.codeplex.com/?p=vacid}

\item
Verfified Software Competition \cite{fil:pas:stu:12,kle:mul:sha:lea:wus:alk:art:bro:cha:coh:hil:jac:lei:mon:pie:pol:rid:sma:tob:tue:ulb:wei:11}. 

\item
VerifyThis Competition: \\
\url{http://www.pm.inf.ethz.ch/research/verifythis.html}

\end{itemize}
The main difference from the benchmark
presented here is that they do not target C explicitly.
Yet, they sometimes use C code in their statements, and sometimes
participants are using C for their solutions.

There are also benchmarks that \emph{are} about specific programming languages,
which often have a C section, like for example:
\begin{itemize}
\item
Collection of Verification Tasks of the SoSy-Lab at the Ludwig-Maximilian University in Munich: \\
\url{https://github.com/sosy-lab/sv-benchmarks/tree/master/c}

\end{itemize}
However, in that case the focus is not so much on formally proving that programs
are fully correct, but on benchmarking the performance of non-interactive tools for discovering problems in the programs.

\section*{Acknowledgments}
Thanks to Jean-Christophe Filli\^atre for many helpful comments.

\bibliographystyle{plain}
\bibliography{cbench}

\end{document}